\documentclass[conference]{IEEEtran}

\usepackage{calc}
\usepackage{stfloats}
\usepackage{calc}
\usepackage{microtype}
\usepackage[intlimits,sumlimits,namelimits]{amsmath} 
\everymath{\displaystyle} 
\usepackage{amssymb}
\usepackage{amsfonts}
\usepackage{psfrag}
\usepackage{verbatim}
\usepackage{multirow}
\usepackage{multicol}
\usepackage{algorithm2e}
\usepackage{blkarray}
\usepackage{wrapfig}
\usepackage{paralist}
\usepackage{trfsigns}
\usepackage{version}
\usepackage{wasysym}
\usepackage{bm}
\usepackage{color}
\usepackage{xcolor}
\usepackage{graphicx}
\usepackage[most]{tcolorbox}
\usepackage[export]{adjustbox} 
\usepackage{caption}
\usepackage[
    backend=biber,
    style=numeric,
    doi=false,
    isbn=false,
    url=false,
    eprint=false
]{biblatex}
\renewbibmacro{in:}{}
\appto\bibsetup{\flushbottom}
\usepackage{flushend}
\usepackage{catchfile}
\usepackage{pgffor}
\usepackage{mathtools}
\usepackage{bigints}
\usepackage{xfrac}
\usepackage{ifthen}

\def\vec#1{\underline{#1}}
\def\mat#1{{\mathbf #1}}

\def\1_2{{\frac{1}{2}}}

\def\rv#1{\boldsymbol{#1}}

\DeclareMathOperator{\E}{E}

\def\ge{\geqslant}

\def\d{\mathrm{d}}

\def\NewR{{\rm I\hspace{-.17em}R}}

\def\Eq#1{(\ref{#1})}

\def\Sec#1{Sec.~\ref{#1}}
\def\SubSec#1{Subsec.~\ref{#1}}

\def\Fig#1{Fig.~\ref{#1}}

\def\Appendix#1{Appendix~\ref{#1}}

\def\Example#1{Example~\ref{#1}}

\def\Box#1{{\fboxsep2pt\fbox{$#1$}}}

\newlength\EqLen

\def\ScaleInner#1{%
  \settowidth{\EqLen}{#1}
  \ifdim\EqLen < \columnwidth%
    \begin{equation*}%
      \begin{minipage}{\EqLen}#1\end{minipage}%
    \end{equation*}%
  \else%
    \begin{equation*}%
      \resizebox{0.99\columnwidth}{!}{\begin{minipage}{\EqLen}#1\end{minipage}}%
    \end{equation*}%
  \fi%
}%

\def\Scale#1
{
  \ScaleInner{$ #1 $}
}

\def\LongVersion#1{}

\def\citep#1{(\cite{#1})}

\usepackage{amsthm}



\newtheorem{examplex}{Example}[section]
\newenvironment{example}
{\pushQED{\qed}\examplex}
{\popQED\endexamplex}

\makeatletter
\newcommand\SaveEquation[2]{\@namedef{equation@#1}{#2}}
\newcommand\UseEquation[1]{\@nameuse{equation@#1}}
\makeatother

\SetKw{KwBy}{by}

\SetCommentSty{mycommfont}

\newcommand\mystrut{\rule[-1pt]{0pt}{.8em}}
\newtcbox{\intextTCB}{on line, boxrule=0pt, boxsep=0pt, top=2pt,
  left=2pt, bottom=2pt, right=2pt, colback=blue!20, colframe=white,
  fontupper={\mystrut}}
\newcommand{\custommark}[1]{\textsuperscript{\intextTCB{\scriptsize\normalfont #1}}}
\usepackage{enotez}
\DeclareInstance{enotez-list}{custom}{paragraph}{
  heading=\section*{#1},
  notes-sep=0.5\baselineskip,
  format=\normalsize\normalfont\leftskip3em,
  number=\makebox[0pt][r]{\intextTCB{\normalsize\normalfont#1:}\hskip0.5em}\ignorespaces,
}
\setenotez{
  list-name=Endnotes,
  backref,
  mark-format = {\custommark},
  mark-cs = {}
}
\let\footnote=\endnote

\newtcolorbox{YellowBox}{
  enhanced,
  boxrule=0pt,frame hidden,
  borderline east={1mm}{0pt}{yellow!90!black},
  borderline west={1mm}{0pt}{yellow!90!black},
  colback=yellow!40!white,
  sharp corners,
  grow sidewards by=1.5mm,
  top=0mm,
  bottom=0mm,
  left*=0mm,
  right*=0mm
}

\date{}

\author{\IEEEauthorblockN{\textbf{Uwe D.\ Hanebeck}}\\
\IEEEauthorblockA{Intelligent Sensor-Actuator-Systems Laboratory (ISAS)\\
Institute for Anthropomatics and Robotics\\
Karlsruhe Institute of Technology (KIT), Germany\\
Uwe.Hanebeck@kit.edu}}

\title{Closed-Form Information-Theoretic\\Roughness Measures for Mixture Densities}

\usepackage[acronyms,nopostdot,nonumberlist,nogroupskip]{glossaries}

\makeglossaries

\newcommand*{\Acro}[4][]{%
    \ifthenelse { \equal {#1} {} }%
    { \newacronym{#2}{#3}{#4} }%
    { \newacronym[#1]{#2}{#3}{#4} }%
}

\Acro{FI}{FI}{Fisher Information}
\Acro{GRM}{GRM}{Gaussian Root Mixture}
\Acro{GM}{GM}{Gaussian Mixture}
\Acro[\glslongpluralkey={mixture densities}]{MD}{MD}{mixture density}
\Acro{pdf}{pdf}{probability density function}
\Acro[\glslongpluralkey={root densities}]{RD}{RD}{root density}
\Acro{RM}{RM}{root mixture}

\begin{document}

\maketitle
\thispagestyle{empty}
\pagestyle{empty}


\begin{abstract}
    %
%
We calculate the smoothest mixture density under a variety of prescribed specifications.
%
%
This includes constraints on certain moments, specifications on density values and/or its derivatives, and prescribed probability masses in certain regions.
%
%
As a roughness measure, we use \gls{FI} in the space of mixtures $\cal M$.
%
%
For mixtures, \gls{FI} cannot be calculated in closed form.
%
%
We define the space $\cal R$ of \glspl{RM} living on the Hilbert sphere.
A transformation of \gls{FI} to $\cal R$ admits a closed-form solution and yields the desired result in $\cal M$.
%
%
This naturally leads to a tandem processing with two density representations maintained simultaneously in $\cal R$ and $\cal M$.
\gls{FI} is calculated in \gls{RM} space $\cal R$ while the constraints are evaluated in mixture space $\cal M$.

\medskip

\renewcommand\IEEEkeywordsname{Keywords}
\begin{IEEEkeywords}
    Mixture density, Gaussian mixture, information measure, Fisher information, smooth density, roughness measure, square-root densities, closed-form expression.
\end{IEEEkeywords}
\end{abstract}


\printglossary[title=Glossary of Terms,type=\acronymtype]

\section{Introduction} \label{Sec_Intro}

\subsection{Context}

%
%
In many applications we are given some specifications on a \gls{pdf}
and want to find a \gls{pdf} that adds as little information as possible to what is specified.
%
%
Intuitively, this corresponds to the smoothest \gls{pdf} under the specifications.
%
%
The \gls{pdf} is selected from a certain class of parametric or non-parametric \glspl{pdf}.
%
%
The goal is to find the smoothest \gls{pdf} from this class in an information theoretic sense meeting the specifications.

%
%
Simple specifications include prescribing certain moments or confining the \gls{pdf} to be nonzero on certain intervals and zero in others.
%
%
More complex specifications assume certain realization/density pairs $(\vec{x}_i, f(\vec{x}_i))$, $i=1 \mathord{:} N$%
\footnote{We use the notation $i \mathord{:} j$ to denote integer sequence $\{i,i+1, \ldots, j-1, j\}$ from~\cite{golubMatrixComputations3rd1996}.}
or corresponding tolerance bands.
%
%
Typically, the constraints on the \gls{pdf} resulting from the specifications do not fully characterize the \gls{pdf}.
Enforcing smoothness can be viewed as a \emph{regularizer} for underdetermined optimization.

\begin{figure}
    \begin{center}
        \includegraphics{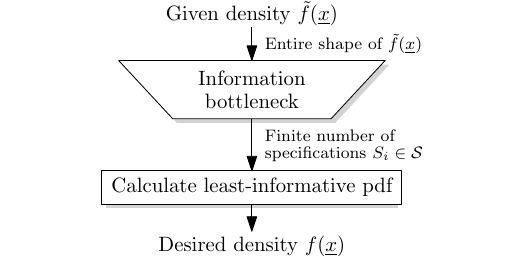}
    \end{center}
    \caption{%
        A finite number of specifications, say, moments or samples, are extracted from the shape of the given density $\tilde{f}(\vec{x})$.
        Compared to the entire shape of $\tilde{f}(\vec{x})$, the set of specifications ${\cal S}$ contains less information.
        The extraction of specification comprises an information bottleneck.
    }
    \label{Fig_bottleNeck}
\end{figure}

%
%
W.l.o.g.\ we can assume the existence of an \emph{unknown} underlying \gls{pdf} $\tilde{f}(\vec{x})$
%
%
with $\vec{x} \in \NewR^D$, $D$ the number of dimensions, of which we only have a set ${\cal S}$ of specifications, see \Fig{Fig_bottleNeck}.
%
%
The set $\cal S$ is not sufficient to fully describe  $\tilde{f}(\vec{x})$.
%
%
When reconstructing an approximation $f(\vec{x})$ of  $\tilde{f}(\vec{x})$, we do not want to artificially add unwarranted information.%
\footnote{By selecting the density $f(\vec{x})$ from a certain class of densities, we inadvertently add information.}
We desire the least-informative \gls{pdf} given the specification.

\subsection{Application Examples}

%
%
Some selected application examples include
(1)~increasing the number of parameters of density representation, e.g., increasing the number of components of a mixture \gls{pdf},
(2)~replacing a moment representation of $\tilde{f}(\vec{x})$ with a continuous density representation $f(\vec{x})$,
and (3)~estimating a continuous \gls{pdf} representing a given set of samples.
%
%
Common to these examples is the insufficient specification of the desired \gls{pdf}, which would yield degenerate results.
%
%
For example, in the density estimation case, a kernel density estimator with Gaussian kernels would degenerate in the sense of zero kernel widths.

\section{Problem Formulation} \label{Sec_ProbForm}

%
%
First, we are given a set of specifications $S_i(f) = 0 \,,\; S_i \in {\cal S}$ about the desired \gls{pdf} $f(\vec{x})$.
%
%
We assume an underlying \gls{pdf} $\tilde{f}(\vec{x})$ that is unknown, see \Fig{Fig_bottleNeck},
that we would like to approximate with the desired \gls{pdf} $f(\vec{x})$.%
\footnote{The underlying \gls{pdf} $\tilde{f}(\vec{x})$ is mainly a vehicle for explanation.
    However, in some cases, the goal might indeed be to reconstruct $\tilde{f}(\vec{x})$ from given moments or samples.
    It is important to note that $\tilde{f}(\vec{x})$ is completely unknown and not used anywhere in the procedure of calculating the desired \gls{pdf} $f(\vec{x})$.}
%
%
Specifications about $\tilde{f}(\vec{x})$ can be of the form
(1)~moments or central moments, e.g., mean, covariance,
(2)~given density values $\tilde{f}(\vec{x}_i)$ at selected $\vec{x}_i$ and/or derivatives,
(3)~probabilities of $\rv{\vec{x}}$ being in certain regions,
and (4)~given samples of $\tilde{f}$.
%
%
We also allow ranges or tolerance bands%
\footnote{Ranges or tolerance bands could be given for moments, values, or even samples.
    An example would be a mean constraint of the form $E\{\rv{x}\} \in [-0.2,0.2]$.
    Another example is a value constraint of the form $f(x_i) \in [\underline{y},\overline{y}]$.}
for certain specifications.

%
%
Second, we are given a certain structure of the \gls{pdf} $f(\vec{x})$.
%
%
This can be achieved by confining $f(\vec{x})$ to a certain class of densities, say, spherically invariant densities \cite{rangaswamyNonGaussianRandomVector1993}.
%
%
More concrete, the \gls{pdf} could be fixed to being of a specific type, say, a Gaussian density.
%
%
Here, our focus is on \glspl{pdf} from the class of \glspl{MD} with a given number $L$ of components
\begin{equation}
    f(\vec{x}) = \sum_{i=1}^L w_i f_i(\vec{x})
    \label{Eq_GenericMixture}
\end{equation}
with positive weights $w_i > 0$, $i \in 1 \mathord{:} L$, summing up to one
\begin{equation}
    \sum_{i=1}^L w_i = 1 \enspace ,
\end{equation}
and mixture component \glspl{pdf} $f_i(\cdot)$, $i \in 1 \mathord{:} L$.
%
%
As a concrete density type, we will consider Gaussian components.

%
%
Our desired outcome is a \gls{MD} $f(\vec{x})$ that does not add any artificial constraints beyond the specifications given.
More specific, we require appropriate mixture weights $w_i>0$, $i \in 1 \mathord{:} L$, and parameters of the component pdfs $f_i(\cdot)$, $i \in 1 \mathord{:} L$.

%
%
For solving this problem, we require a measure of the information content of pdf $f(\vec{x})$ or equivalently of its smoothness/roughness.
This measure is then used to find the mixture $f(\vec{x})$ with the least amount of information (or equivalently the smoothest \gls{pdf}) given the specifications ${\cal S}$.

%
%
In the next section, we review common smoothness and information measures and their suitability for the task at hand.

\section{State-of-the-Art} \label{Sec_State_of_Art}

%
%
In this section, we briefly review the state-of-the-art in calculating least-informative \glspl{pdf}.
%
%
Given a set of specifications on the \gls{pdf}, a least-informative \gls{pdf} does not add further information.
%
%
A \gls{pdf} with little information content intuitively exhibits a high level of smoothness or equivalently a low level of roughness.
%
%
We will focus on methods based on curvature, entropy, and \gls{FI} as a basis for the derivation of the proposed new method in \Sec{Sec_ClosedForm}.

%
%
\subsection{Mean Curvature}
%
%
At first look, mean curvature would be a good candidate for quantifying smoothness.
%
%
The problem is that integrating over the curvature is numerically difficult.
%
%
Hence, curvature is often approximated by the second derivative of the considered \gls{pdf}.
%
%
Expressions of this type are used as roughness penalties for smoothing splines \cite[177]{reinschSmoothingSplineFunctions1967}.

%
%
For simplicity, we only consider the scalar case, which is already quite complex.
%
%
The local curvature of $f$ is defined as%
\footnote{Defining curvature in the multivariate case is much more complicated, see for example \cite{leeRiemannianManifoldsIntroduction1997}.}
\begin{equation}
    k(f(x)) = \frac{f''(x)}{\left( 1 + \left( f'(x) \right)^2 \right)^{\sfrac{3}{2}}} \enspace .
    \label{Eq_curvature1D}
\end{equation}
%
%
The mean squared curvature is given as the integral over the local curvature as
\begin{equation}
    k(f) = \int_{\NewR} k^2 \big( f(x) \big) \, \d x \enspace,
    \label{Eq_mean_curvature1D}
\end{equation}
which cannot be solved in closed form for densities of interest here.
%
%
For this reason, the local curvature is often approximated with the second derivative of $f$ as $k(f(x)) = f''(x)$.
%
%
In higher dimensions, this can be generalized to $k(f(\vec{x})) = \nabla^2 f(\vec{x})$ with Hessian operator $\nabla^2 = \nabla \cdot \nabla^\top$.
%
%
The resulting mean squared curvature
\begin{equation}
    k(f) = \int_{\NewR^D} \left| \nabla^2 f(\vec{x}) \right|^2 \, \d \vec{x}
    \label{Eq_mean_curvature_Hessian}
\end{equation}
with Frobenius norm $|\cdot|$ can often be calculated in closed form, e.g., for Gaussian or \gls{GM} densities $f(\vec{x})$.
%
%
However, although this simplified mean curvature might be useful for, e.g., spline smoothing, it is not a good indicator of smoothness for \glspl{pdf}.
%
%
The reason is that in a typical reference situation, i.e., minimization of the roughness of a density with zero mean and unit variance,
minimization of the curvature does not yield a Gaussian density as would be expected.
On the other hand, maximization of the entropy or minimization of the \gls{FI} yields the expected Gaussian density in this case as we will see in \SubSec{SubSec_Entropy} and \SubSec{SubSec_FI}.

\noindent
{\bfseries Summary:}
%
%
The mean curvature of a general density according to \Eq{Eq_curvature1D} and \Eq{Eq_mean_curvature1D} are difficult to compute even in the univariate case.
This is exacerbated for mixture densities.
For this reason, the curvature is often approximated by the second derivative of $f(x)$ in 1D.
In the multivariate case, the definition of curvature is even more complicated, which is another reason for approximation by the Hessian of $f(\vec{x})$.
%
%
Unfortunately, the simplified mean curvature, although easily computable for mixtures, does not yield the expected results in simple reference cases.

\subsection{Entropy} \label{SubSec_Entropy}
%
%
Most methods for finding the least-informative \gls{pdf} under given specifications use the principle of maximum entropy.
%
%
It was conceived by Jaynes in 1957 \cite{jaynesInformationTheoryStatistical1957,jaynesInformationTheoryStatistical1957a}.
%
%
It received its share of criticism \cite{friedmanJaynesMaximumEntropy1971},
%
%
but was eventually adopted by the research community.
%
%
One example is the use of maximum entropy for the trigonometric moment problem and orthogonal polynomials \cite{landauMaximumEntropyMoment1987}.

For a continuous density $f(\vec{x})$, the so-called differential entropy is defined as
\begin{equation}
    \E\{ -\log\left( f(\vec{x}) \right) \}
    = - \smashoperator{\int_{\vec{x} \in \NewR^D}} f(\vec{x}) \log\left( f(\vec{x}) \right) \d \vec{x} \enspace .
\end{equation}
%
%
The differential entropy has many nice properties.
%
%
However, it is an ad-hoc generalization of the famous Shannon entropy for discrete random variables to the continuous setting.
It can assume negative values and is not invariant under a change of variables.

%
%
A table with entropy expressions for standard \glspl{pdf} is provided in \cite{lazoEntropyContinuousProbability1978}.
%
%
Expressions for the entropy in multivariate settings are given in \cite{darbellayEntropyExpressionsMultivariate2000}, but this does not include mixture distributions.
%
%
Because of the logarithm in the expression of the differential entropy used in maximum entropy methods, computation is complicated beyond simple cases such as Gaussian densities.
%
%
This is especially a problem for mixture densities as these lead to logarithms of sums.
%
%
As a result, entropy is often calculated via numerical integration, e.g., Monte Carlo, which is especially complex in multivariate settings.

%
%
This led to an extensive development of approximations and bounds, especially for mixture densities.
%
%
Differential entropy for \glspl{GM} is approximated in \cite{MFI08_HuberBailey} by a Taylor-series expansion of the logarithm of the \gls{GM} around each component mean.
For large component variances, splitting of components is required to maintain a desired accuracy.
In \cite{arXiv14_Hanebeck}, a deterministic sample representation is approximated by a piecewise constant density to facilitate the computation of the relative entropy.
A piecewise constant approximation of a \gls{GM} is proposed in \cite{zhangApproximatingDifferentialEntropy2017}.
%
%
Sharp bounds on the so-called entropy concavity deficit are derived in \cite{melbourneDifferentialEntropyMixtures2022}, i.e., the difference between the mixture entropy and the sum of component entropies.
In \cite{moshksarArbitrarilyTightBounds2016}, lower and upper bounds on the differential entropy for the special case of \glspl{GM} are provided when all components have identical variances and only differ in their means and weights.
For a symmetric \gls{GM} with two components with equal weights and equal variances, lower and upper bounds on the differential entropy are given in \cite{michalowiczCalculationDifferentialEntropy2008}.

\noindent
{\bfseries Summary:}
%
%
As it requires integration over a logarithm of $f(\vec{x})$, differential entropy can be calculated in closed form only in rare cases.
In particular, \glspl{MD} do not admit a closed-form solution.
In that case, one has to rely on the approximations or bounds mentioned above.

\subsection{Fisher Information} \label{SubSec_FI}

%
%
Using \gls{FI} for finding the smoothest continuous density has first been proposed in \cite{goodNonparametricRoughnessPenalty1971}.
It was used as a roughness penalty in maximum likelihood density estimation \cite{goodNonparametricRoughnessPenalties1971} based on orthogonal Hermite polynomials.
%
%
Similarly, \gls{FI} is employed in \cite{vannucciPreventingDiracDisaster1997} for wavelet-based density estimation.
As a roughness measure, it prevents the density estimate to come too close to the Dirac functions representing the observations.
%
%
In \cite{huberRobustEstimationLocation1964}, \gls{FI} is used for the robust estimation of a location parameter.
%
%
This method has been extended to minimizing \gls{FI} over mixtures in \cite{bickelMinimizingFisherInformation1983}.

%
%
For deriving \gls{FI}, we start with the so-called score \cite[18]{toscaniScoreFunctionsGeneralized2017} given by
\begin{equation}
    s(\vec{x}; \vec{\theta}) = \frac{\partial}{\partial \vec{\theta}} \log\left( f(\vec{x}; \vec{\theta}) \right)
    \label{Eq_ScorePar}
\end{equation}
for some density $f$ depending on a vector parameter $\vec{\theta}$ or equivalently
\begin{equation}
    s(\vec{x}; \vec{\theta}) = \frac{\partial f(\vec{x}; \vec{\theta})}{\partial \vec{\theta}} \frac{1}{f(\vec{x}; \vec{\theta})}
    = \frac{\nabla f(\vec{x}; \vec{\theta})}{f(\vec{x}; \vec{\theta})} \enspace .
    \label{Eq_ScoreParDiv}
\end{equation}
%

%
%
Following \cite[29]{goodNonparametricRoughnessPenalty1971}, we define roughness as the difference between the original density
$f(\vec{x})$ and a copy $f(\vec{x}+\vec{\theta})$ shifted by a small displacement $\theta$.
Thus, we have $f(\vec{x}; \vec{\theta}) = f(\vec{x} + \vec{\theta})$ and the score is
\begin{equation}
    s(\vec{x}; \vec{\theta}) = \frac{\partial f(\vec{x} + \vec{\theta})}{\partial \vec{\theta}} \frac{1}{f(\vec{x} + \vec{\theta})}
    = \frac{\partial f(\vec{x} + \vec{\theta})}{\partial \vec{x}} \frac{1}{f(\vec{x} + \vec{\theta})} \enspace .
\end{equation}
With $\vec{\theta} \rightarrow \vec{0}$, the score \Eq{Eq_ScoreParDiv} can be written as
\cite[2]{nagyFisherInformationDensity2022}
\begin{equation}
    s(\vec{x}) = \frac{\partial f(\vec{x})}{\partial \vec{x}} \frac{1}{f(\vec{x})}
    = \frac{\nabla f(\vec{x})}{f(\vec{x})}
    \label{Eq_Score}
\end{equation}
or the score \Eq{Eq_ScorePar} as
\begin{equation}
    s(\vec{x}) = \frac{\partial}{\partial \vec{x}} \log\left( f(\vec{x}) \right) \enspace .
    \label{Eq_Score_log}
\end{equation}
Integrating over the entire domain in the original space of mixtures ${\cal M}$ gives the \gls{FI} \cite[p.~15, (1)]{toscaniScoreFunctionsGeneralized2017}
\begin{equation}
    I_F^{\cal M}(f) = \smashoperator{\bigintsss_{\{ \NewR^D,\, f>0 \}}} \left| \frac{\nabla f(\vec{x})}{f(\vec{x})} \right|^2 f(\vec{x}) \, d \vec{x}
    \label{Eq_FI}
\end{equation}
or alternatively
\begin{equation}
    I_F^{\cal M}(f) = \smashoperator{\bigintsss_{\{ \NewR^D,\, f>0 \}}} \frac{\left| \nabla f(\vec{x}) \right|^2}{f(\vec{x})} \, d \vec{x} \enspace .
    \label{Eq_FI_simplified}
\end{equation}
We exclude regions where the density approaches zero.

%
%
The \gls{FI} is often simpler to compute than the differential entropy.
However, the division by $f(\vec{x})$ in \Eq{Eq_FI_simplified} still makes computation difficult especially for mixture densities.
%
%
In \cite{bickelMinimizingFisherInformation1983}, although minimizing \gls{FI} over mixtures has been considered, computational issues were not addressed.

\noindent
{\bfseries Summary:}
\gls{FI} in the form \Eq{Eq_FI} or \Eq{Eq_FI_simplified} cannot be calculated in closed form for \glspl{MD} because of the division by $f(\vec{x})$.



\section{Closed-form \acrlong{FI} for Mixtures} \label{Sec_ClosedForm}

\subsection{Key Idea}
\begin{figure}
    \begin{center}
        \includegraphics{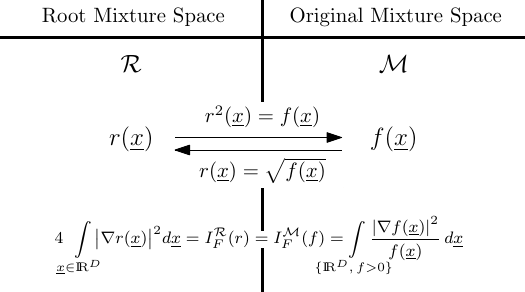}
    \end{center}
    \caption{%
        The two spaces used for calculating least-informative \glspl{pdf} fulfilling given specifications.
    }
    \label{Fig_twoSpaces}
\end{figure}

%
%
The key idea for obtaining a closed-form expression for the \gls{FI} of a mixture density is to work in the space of square-root densities, see \Fig{Fig_twoSpaces}.
%
%
In particular, we define \glspl{RM} that yield the original mixture upon squaring.
%
%
The desired \gls{FI} of the original mixture can be calculated in closed form in the \gls{RM} space.
%
%
We propose to simultaneously maintain both density representations and perform tandem processing in both spaces.
The \gls{FI} is calculated in the \gls{RM} space ${\cal R}$.
The specifications on the \gls{pdf} $f(\vec{x})$ are calculated in the original mixture space ${\cal M}$.

\subsection{Root Densities}

%
%
Square roots of densities $r(\vec{x})=\sqrt{f(\vec{x})}$ have already been considered in \cite{goodNonparametricRoughnessPenalties1971}.
The reason for doing so was to ensure that the squared density $f(\vec{x}) = r^2(\vec{x})$ is non-negative.
%
%
For the same reason, densities based on squared Fourier series are introduced in \cite{MFI06_Brunn-Fourier},
generalized for multivariate densities in \cite{CDC06_Brunn-Fourier}, and improved in \cite{JAIF16_Pfaff}.

%
%
Here, we use square-root \glspl{pdf}, or in short \glspl{RD}, for an entirely different reason.
Our considered density representations, e.g., Gaussian mixtures, are non-negative anyway.
Instead, we want to obtain a closed-form expression for the \gls{FI} in \Eq{Eq_FI_simplified}.

%
%
The unit-mass constraint for the \gls{pdf} $f(\vec{x})$
\begin{equation}
    \int_{\NewR^D} f(\vec{x}) \, \d \vec{x} = 1
\end{equation}
is equivalent to the constraint
\begin{equation}
    \int_{\NewR^D} r^2(\vec{x}) \, \d \vec{x} = 1
\end{equation}
for the \gls{RD} $r(\vec{x})$.
This means that $r(\vec{x})$ is restricted to the infinite-dimensional unit sphere $S^{\infty}$.
This manifold is also called the Hilbert sphere \cite[2]{daiStatisticalInferenceHilbert2021}.
%
%
When we restrict the \glspl{RD} $r(\vec{x})$ to be non-negative, i.e., $r(\vec{x}) \ge 0$ $\forall \vec{x}$,
$r(\vec{x})$ is restricted to the ``positive orthant'' of the Hilbert sphere.

\subsection{\acrlong{FI} for Root Densities}
%
%
For \glspl{RD}, the expression for the \gls{FI} can be simplified.
In the univariate case, $D=1$, with
\begin{equation}
    r'(\vec{x}) = \frac{\d}{\d\vec{x}} r(\vec{x})
    = \frac{\d}{\d\vec{x}} \sqrt{f(\vec{x})}
    = -\frac{f'(\vec{x})}{2\sqrt{f(\vec{x})}} \enspace ,
\end{equation}
we have
\begin{equation}
    \big( r'(\vec{x}) \big)^2 = \frac{\big( f'(\vec{x}) \big)^2}{4 \, f(\vec{x})} \enspace .
\end{equation}
Finally, we can rewrite the \gls{FI} in the original density space in \Eq{Eq_FI_simplified} in the \gls{RD} space ${\cal M}$ as
\begin{equation}
    I_F^{\cal R}(r) = 4 \smashoperator{\int_{\vec{x} \in \NewR^D}} \big( r'(\vec{x}) \big)^2 \d \vec{x}
    \text{ with } r = \sqrt{f} \enspace .
    \label{Eq_FI_root}
\end{equation}
%
%
%
In the multivariate case, we have
\begin{equation}
    I_F^{\cal R}(r) = 4 \smashoperator{\int_{\vec{x} \in \NewR^D}} \big| \nabla r(\vec{x}) \big|^2 d \vec{x}
    \text{ with } r = \sqrt{f} \enspace .
    \label{Eq_FI_root_multivariate}
\end{equation}

%
%
These expressions for $I_F^{\cal R}(r)$ in the \gls{RD} space ${\cal R}$ can be calculated in closed form, e.g., for \glspl{RM}.
%
%
In fact, they correspond to the simplified expressions for mean curvature in \Eq{Eq_mean_curvature_Hessian}
that are, however, given in the original density space.

\subsection{Root Mixtures}
%
%
We define \glspl{RD} in the case of \glspl{MD} of the form \Eq{Eq_GenericMixture} as
\begin{equation}
    r(\vec{x}) = \sum_{i=1}^R v_i \, r_i(\vec{x}) \enspace .
    \label{Eq_GenericRootMixture}
\end{equation}
These square-root \glspl{MD} will be abbreviated as \glspl{RM}.

%
%
\subsubsection{Conversion of Root Mixture to Mixture}
The conversion of a \gls{RM} in \Eq{Eq_GenericRootMixture} to a mixture in \Eq{Eq_GenericMixture} is unique, always exists, and done in closed form according to
\begin{equation}
    \begin{aligned}
        f(\vec{x}) & = r^2(\vec{x})                                                                                 \\
                   & = \left(\sum_{i=1}^R v_i \, r_i(\vec{x})\right)  \left(\sum_{i=1}^R v_i \, r_i(\vec{x})\right) \\
                   & = \sum_{i=1}^R \sum_{j=1}^R v_i \, v_j \, r_i(\vec{x}) \, r_j(\vec{x}) \enspace .
    \end{aligned}
\end{equation}
%
%
This expression contains redundant terms as $r_i(\vec{x}) \, r_j(\vec{x}) = r_j(\vec{x}) \, r_i(\vec{x})$ for $i \ne j$ and can be written as
\begin{equation}
    f(\vec{x}) = \sum_{i=1}^R v_i^2 \, r_i^2(\vec{x})
    + 2 \, \sum_{i=1}^R \sum_{j=i+1}^R v_i \, v_j \, r_i(\vec{x}) \, r_j(\vec{x})
\end{equation}
or%
\footnote{Please note that the summations range in $i \in [1,R]$ and $j \in [i,R]$ as we exploit symmetry.}
\begin{equation}
    f(\vec{x}) = \sum_{i=1}^R \sum_{j=i}^R v_i \, v_j \, c_{i,j} \, r_i(\vec{x}) \, r_j(\vec{x})
    \label{Eq_RM_2_GM}
\end{equation}
with
\begin{equation}
    c_{i,j} = \begin{cases} 1 \,,\; i=j \\ 2 \,,\; i \ne j \end{cases} \enspace .
\end{equation}
The \gls{MD} $f(\vec{x})$ in \Eq{Eq_RM_2_GM} contains a total of
\begin{equation}
    L = R \cdot (R+1) / 2
\end{equation}
components%
\footnote{When we prespecify the minimum number of components $L$ we desire of a \gls{GM} in the original mixture space ${\cal M}$,
    the number of \gls{RM} components in the \gls{RM} space ${\cal R}$ for a full expansions according to \Eq{Eq_RM_2_GM} are given by
    \begin{equation}
        R = \left\lceil \frac{\sqrt{8 \cdot L + 1} - 1}{2} \right\rceil \enspace ,
    \end{equation}
    where $\lceil \cdot \rceil$ denotes the next largest integer.
}
%
%
and is non-negative by definition (although its weights may be negative).
However, without additional constraints on $r(\vec{x})$, the unit mass constraint may be violated.

%
%
For deriving specific constraints, we need to consider specific mixture densities.
Here, we will consider mixtures of the form \Eq{Eq_GenericMixture} with the special case of Gaussian components
    {\small
        \begin{equation}
            f_i(\vec{x}) = \frac{1}{\sqrt{|2 \pi \mat{\Sigma}_i|}}
            \exp\left\{ - \frac{1}{2} (\vec{x}-\vec{x}_i)^\top \cdot \mat{\Sigma}_i^{-1} \cdot (\vec{x}-\vec{x}_i) \right\}
            \label{Eq_GM_components}
        \end{equation}
    }
\noindent
with $\vec{x} \in \NewR^D$ abbreviated as%
\footnote{%
    For a multivariate normal random vector $\rv{\vec{x}} \in \NewR^D$ we denote a Gaussian density by
    \begin{equation}
        f(\vec{x}) = N(\vec{x}; \vec{m}, \mat{C})
    \end{equation}
    with realization $\vec{x} \in \NewR^D$, mean vector $\vec{m} \in \NewR^D$, and covariance matrix $\mat{C}$.
}
\begin{equation}
    f_i(\vec{x}) = N(\vec{x}; \vec{x}_i, \mat{\Sigma}_i) \enspace .
\end{equation}


%
%
The corresponding Gaussian \gls{RM} for $\vec{x} \in \NewR^D$ is given by \Eq{Eq_GenericRootMixture} with Gaussian components
    {\small
        \begin{equation}
            r_i(\vec{x}) = \frac{1}{\sqrt{|2 \pi \mat{P}_i|}}
            \exp\left\{ - \frac{1}{2} (\vec{x}-\vec{\rho}_i)^\top \cdot \mat{P}_i^{-1} \cdot (\vec{x}-\vec{\rho}_i) \right\}
        \end{equation}
    }
\noindent
with weights $v_i$, mean vectors $\vec{\rho}_i$, and covariance matrices $\mat{P}_i$.

%
%
According to \Eq{Eq_RM_2_GM}, the conversion of a Gaussian \gls{RM} with $R$ components and parameters $v_i$, $\vec{\rho}_i$, $\mat{P}_i$
to a \gls{GM} results in $L = R\cdot(R+1)/2$ components with $w_i$, $\vec{x}_i$, $\mat{\Sigma}_i$ given in \Appendix{Sec_Appendix}.
%
%
Please note that the conversion is especially simple for Gaussian components as the product of Gaussians is again Gaussian.

%
%
\paragraph*{Basic Constraints on Parameters of $r(\vec{x})$}
The \gls{GM}  $f(\vec{x})$ resulting from the conversion of the Gaussian \gls{RM} $r(\vec{x})$ does not necessarily have positive weights that sum to one.
For this reason, basic constraints on the parameters of $r(\vec{x})$ in \gls{RM} space ${\cal R}$ are derived from the constraints
\begin{equation}
    w_i>0 \,,\; i = 1 \mathord{:} L \,,\; \sum_{i=1}^L w_i = 1
\end{equation}
on the \gls{GM} $f(\vec{x})$ in the original mixture space ${\cal M}$.

%
%
\paragraph*{Number of Parameters}
%
%
The number of parameters of the Gaussian \gls{RM} $r(\vec{x})$ and the \gls{GM} $f(\vec{x})$ are equal as the squaring operation in \Eq{Eq_RM_2_GM} does not add parameters.
%
%
In the multivariate case, the parameters of $r(\vec{x})$  with $R$ components are given by%
\footnote{The covariance matrices $\mat{P}_i$ are positive-definite and symmetric.
    Thus, each $\mat{P}_i$ is specified by $D \cdot (D+1) / 2$ parameters.}
\begin{equation}
    \underbrace{v_1, v_2, \cdots, v_R}_{R}
    \,,\; \underbrace{\vec{\rho}_1, \vec{\rho}_2, \cdots, \vec{\rho}_R}_{R \cdot D}
    \,,\; \underbrace{\mat{P}_1, \mat{P}_2, \cdots, \mat{P}_R}_{R \cdot \frac{D \cdot (D+1)}{2}} \enspace .
\end{equation}
With the above mentioned basic constraint we lose one degree of freedom in the weights, so that we end up with
\begin{equation}
    (R-1) + R \cdot D + R \cdot \frac{D \cdot (D+1)}{2} = R \cdot \frac{D^2 + 3 D + 2}{2} - 1
\end{equation}
parameters.

%
%
\paragraph*{Limiting the Number of Components}
The conversion from an \gls{RM} to a mixture becomes impractical for large $R$ as $L$ increases quadratically with $R$.
However, there are several simple and effective ways to alleviate this problem.
%
%
First, we note that there is redundancy in the resulting mixture in \Eq{Eq_RM_2_GM} as the number of parameters of $f(\vec{x})$ stays the same as in $r(\vec{x})$,
but the number of components is larger.
%
%
This redundancy could be exploited in a subsequent component reduction of $f(\vec{x})$.
%
%
However, this is computationally costly.
%
%
A better alternative is to exploit the decreasing overlap between distant mixture components.
For mixtures with finitely supported components, say, triangular mixtures, the overlap becomes zero after a certain distance.
For components with infinite support, e.g., in the case of \glspl{GM}, the overlap quickly decreases in practical settings.
%
%
Hence, not every combination of components has to be multiplied.
This typically results in $L \ll R^2$.

%
%
The extreme case would be to assume no overlaps between mixture components at all.
In this case, which is realistic for \glspl{GM}, the numbers of components would stay exactly the same, i.e., $L = R$.
%
%
A more practical approach is to only consider close neighbors or just the nearest neighbors.
However, closeness does not only depend on the locations of the individual components but also on their extent.
%
%
This calls for an adaptive approach that only includes those pairs of mixture components in the expansion \Eq{Eq_RM_2_GM} that yield new components with significant weights.

%
%
\subsubsection{Conversion of Mixture to Root Mixture}
%
%
An \gls{RM} $r(\vec{x})$ corresponding to a given \gls{MD} $f(\vec{x})$ does not necessarily exist.
This is due to the smaller number of parameters in $r(\vec{x})$ compared to $f(\vec{x})$ so that not every possible $f(\vec{x})$ can be represented by $r(\vec{x})$.
In addition, even when it exists, an appropriate \gls{RM} is non-unique due to, e.g., squared weights.
%
%
Again, we can exploit small overlaps between distant components when performing the conversion.
This reduces the differences in numbers of parameters between the \gls{RM} and the original mixture.

\begin{figure}
    \begin{center}
        \includegraphics[width=\columnwidth]{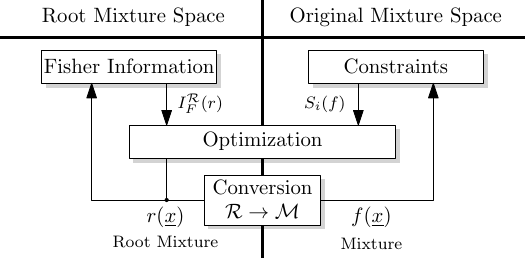}
    \end{center}
    \caption{%
        Overview of optimization via tandem processing in \gls{RM} space space ${\cal R}$ and original mixture space ${\cal M}$.
    }
    \label{Fig_optimization}
\end{figure}

\subsection{Tandem Processing for Optimization}

%
%
The \gls{RM} $r(\vec{x})$ in root mixture space $\cal R$ is convenient as it allows a closed-form calculation of the \gls{FI}.
%
%
In addition, the nonnegativity of the corresponding mixture $f(\vec{x}) = r^2(\vec{x})$ in mixture space $\cal M$ is automatically guaranteed.
%
%
However, the unit integral constraint for $f(\vec{x})$ has to be explicitly ensured by an appropriate constraint in mixture space $\cal M$.
%
%
Also, the specifications on $f(\vec{x})$ are formulated in mixture space $\cal M$.

\begin{figure}
    \def\basePath{Plots/}
    \foreach \i in {1,...,3} {%
            \def\figPath{\basePath compareGauss_\i}
            \begin{center}
                \includegraphics{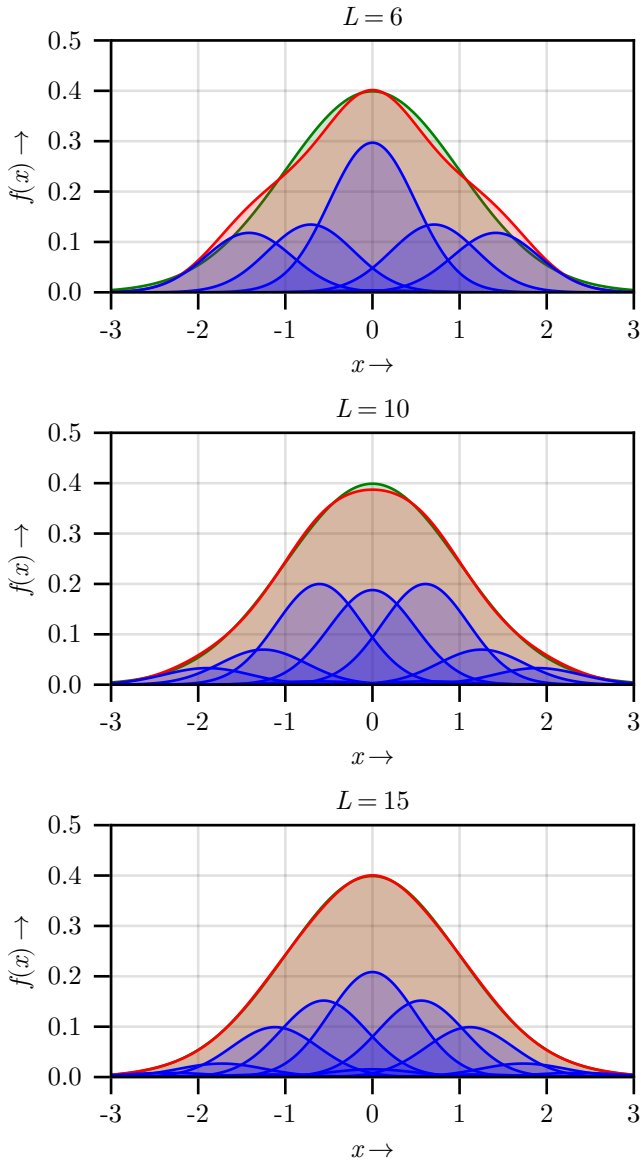}
            \end{center}
        }
    \caption{%
        Red: Smoothest \gls{GM} $f(\vec{x})$ under specifications zero mean and unit variance in \Example{Ex_compareGauss}
        the first example
        for different numbers of components~$L$.
        Blue: Corresponding mixture components $f_i(\vec{x})$, $i \in 1 \mathord{:} L$.
        Green: Gaussian density with zero mean and unit variance as reference.
    }
    \label{Fig_compareGauss}
\end{figure}

%
%
During the optimization procedure, i.e., minimization of \gls{FI}, see \Fig{Fig_optimization}, we maintain both density representations in \gls{RM} space $\cal R$ and in mixture space $\cal M$.
\gls{FI} calculation is done in \gls{RM} space $\cal R$ in tandem with the constraint evaluation in mixture space $\cal M$.

%
%
The optimization problem can be written as
\begin{equation}
    \begin{array}{cl}
        \min_{f \in {\cal M}} & I_F^{\cal M}(f)                              \\[2mm]
        \text{s.t.}           & S_i(f) = 0 \,,\; S_i \in {\cal S} \enspace ,
    \end{array}
\end{equation}
with the \gls{FI} $I_F^{\cal M}(f)$ from \Eq{Eq_FI}.
This is equivalent to
\begin{equation}
    \begin{array}{cl}
        \min_{r \in {\cal R}} & I_F^{\cal R}(r)                              \\[2mm]
        \text{s.t.}           & f = r^2                                      \\
                              & S_i(f) = 0 \,,\; S_i \in {\cal S} \enspace ,
    \end{array}
\end{equation}
with the \gls{FI} $I_F^{\cal R}(r)$ from \Eq{Eq_FI_root} and $I_F^{\cal M}(f) = I_F^{\cal R}(r)$.

\section{Examples} \label{Sec_Examples}

%
%
We will now give two example applications of the proposed new optimization method.
%
%
In the examples, we assume a \gls{GM} $f(\vec{x})$ of the form \Eq{Eq_GenericMixture} with $L$ components of the form
\Eq{Eq_GM_components}.
%
%
We start with a simple example of the smoothest \gls{GM} with zero mean and unit variance, see \Example{Ex_compareGauss}.
In \Example{Ex_Values}, we prescribe density values at certain locations.

%
%
\begin{example}[Zero mean and unit variance]
    \label{Ex_compareGauss}
    %
    %
    Besides the constraints on the weights (larger than zero and unit sum), the specifications are only zero mean and unit variance.
    %
    %
    We know that of all \glspl{pdf} with given covariance matrix, the Gaussian density minimizes the \gls{FI}%
    \footnote{For a given covariance matrix, the Gaussian density also maximizes the relative entropy \cite[p.~411, Example 12.2.1, (12.14)]{coverElementsInformationTheory2006}.}%
    \cite[Lemma~1, p.~184]{parkGaussianAssumptionLeast2013} and \cite{pakAlternativeProofMinimum2018}.
    %
    %
    Hence, we would expect the \gls{GM} $f(\vec{x})$ to approach a Gaussian shape when the number of components grows.
    %
    %
    The results are shown in the original mixture space ${\cal M}$ in \Fig{Fig_compareGauss} for $R \in \{3,4,5\}$ \gls{RM} components.
    With \Eq{Eq_RM_2_GM}, we obtain $L = R \cdot(R+1)/2 \in \{6,10,15\}$ mixture components.
    %
    %
    For $L=6$, the number of parameters in the \gls{GM} $f(\vec{x})$ is not sufficient to approach the expected Gaussian density.
    For $L=10$, $f(\vec{x})$ is closer and for $L=15$, $f(\vec{x})$ is visually indistinguishable from the Gaussian.
    For $R=5$ and $L=15$, it is apparent, that not all components significantly contribute to the density shape as some components are very small.
\end{example}

\begin{figure}
    \def\basePath{Plots/}
    \foreach \i in {1,...,3} {%
            \def\figPath{\basePath values_\i}
            \begin{center}
                \includegraphics{\figPath .pdf}
            \end{center}
        }
    \caption{%
        Red: Smoothest \gls{GM} $f(\vec{x})$ under zero mean and value specifications in \Example{Ex_Values} for different numbers of components~$L$.
        Blue: Corresponding mixture components $f_i(\vec{x})$, $i \in 1 \mathord{:} L$.
    }
    \label{Fig_Values}
\end{figure}

%
%
\begin{example}[Prescribed values and derivatives]
    \label{Ex_Values}
    %
    %
    In this example, we use the weight constraints and zero mean as basic constraints.
    In addition, we specify the following $(x,f(x))$ value pairs: $(-2.0, 0.1)$, $(-1.0, 0.25)$, $(0.0, 0.25)$, and $(1.0, 0.4)$.
    %
    %
    The results are shown in \Fig{Fig_Values} for $R \in \{3,4,5\}$ \gls{RM} components and $L \in \{6,10,15\}$ mixture components.
    %
    %
    For all $L$, the constraints are matched.
    For increasing $L$, the density $f(x)$ becomes smoother.
\end{example}

\section{Conclusions} \label{Sec_Conclude}

%
%
We derive a closed-form expression for the \gls{FI} of a \gls{GM} density $f(\vec{x})$ given in a mixture space ${\cal M}$.
%
%
This is achieved by introducing an \gls{RM} space ${\cal R}$ and calculating the \gls{FI} there.
%
%
The resulting expression consists of a combination of higher-order moments of Gaussian densities, containing nonlinear terms of the component means and covariances.
%
%
We consider very general specifications of the shape, probability mass distribution, and (central) moments of $f(\vec{x})$.
These specifications can also typically be described by nonlinear equations containing component means and covariances.
%
%
Standard optimization procedures for minimization under equality constraints can be employed to find the least-informative \gls{GM} density $f(\vec{x})$ under the given specifications.

%
%

\section{Appendix} \label{Sec_Appendix}


%
%
\paragraph*{Product of Two Multivariate Gaussian Densities}
Given two multivariate Gaussian densities
\begin{equation}
    f_i(\vec{x}) = N(\vec{x}; \vec{x}_i, \mat{\Sigma}_i)
\end{equation}
or
\begin{equation}
    f_i(\vec{x}) = \frac{1}{\sqrt{|2 \pi \mat{\Sigma}_i|}}
    \exp\left\{ -\frac{1}{2} (\vec{x}-\vec{x}_i)^\top \mat{\Sigma}_i^{-1} (\vec{x}-\vec{x}_i) \right\}
\end{equation}
for $i=1,2$, the product $f_3(\vec{x}) = f_1(\vec{x}) \cdot f_2(\vec{x})$ is given by
\begin{equation}
    f_3(\vec{x}) = c_3 \cdot N(\vec{x}; \vec{x}_3, \mat{\Sigma}_3)
\end{equation}
with factor
\begin{equation}
    c_3 = N(\vec{0}; \vec{x}_1 - \vec{x}_2, \mat{\Sigma}_1 + \mat{\Sigma}_2) \enspace ,
\end{equation}
new mean vector
\begin{equation}
    x_3 = \left( \mat{\Sigma}_1^{-1} + \mat{\Sigma}_2^{-1} \right)^{-1}
    \left( \mat{\Sigma}_1^{-1} \vec{x}_1 + \mat{\Sigma}_2^{-1} \vec{x}_2 \right) \enspace ,
\end{equation}
and new covariance matrix
\begin{equation}
    \mat{\Sigma}_3 = \left( \mat{\Sigma}_1^{-1} + \mat{\Sigma}_2^{-1} \right)^{-1} \enspace .
\end{equation}
%
%
Please note that $f_3(\vec{x})$ is an unnormalized density, where the factor $c_3$ reflects the distance between the original means.
Obviously, the new density $f_3(\vec{x})$ as the result of the multiplication of $f_1(\vec{x})$ and $f_2(\vec{x})$ contains less probability mass, the farther $f_1(x)$ and $f_2(x)$ are apart.
The density can easily be normalized by omitting the factor $c_3$.
However, when multiplying two Gaussian mixtures, the factors have to be considered as they take care of the relative weights.
%


\printendnotes[custom]

\printbibliography

\end{document}